\documentclass[preprint,prb,superscriptaddress,amssymb,amsmath, floatfix,citeautoscript]{revtex4}

\usepackage[pdftex]{graphicx}

\newcommand{\w}{\omega}
\renewcommand{\k}{\mathbf{k}}

\newcommand{\Tzt}{T(\mathbf{r},t)}
\newcommand{\vzk}{\mathbf{v}}

\renewcommand{\r}{\mathbf{r}}

\newcommand{\tk}{\tau_{\k}}

\newcommand{\um}{\mu m}

\begin{document}

\title{Multidimensional quasiballistic thermal transport in transient grating spectroscopy}

\author{A. J. Minnich}
\email{aminnich@caltech.edu}
\affiliation{Division of Engineering and Applied Science\\
California Institute of Technology\\
Pasadena, CA 91125}

\begin{abstract}
Transient grating spectroscopy has emerged as a useful technique to study thermal phonon transport because of its ability to perform thermal measurements over length scales comparable to phonon mean free paths (MFPs). While several prior works have performed theoretical studies of quasiballistic heat conduction in transient grating, the analysis methods are either restricted to one spatial dimension or require phenomenological fitting parameters. Here, we analyze quasiballistic transport in a two-dimensional transient grating experiment, in which heat conduction can occur both in- and cross-plane, using an analytic Green's function of the Boltzmann equation we recently reported that is free of fitting parameters. We demonstrate a method by which phonon MFPs can be extracted from these measurements, thereby extending the MFP spectroscopy technique using transient grating to opaque bulk materials.

\end{abstract}

\maketitle

\section{Introduction}

Transient grating spectroscopy (TG) is a noncontact, optical pump-probe experiment that is capable of measuring the in-plane thermal conductivity of thin films and bulk materials \cite{maznev_optical_1998, johnson_phase-controlled_2012}. In this experiment, the sample is impulsively heated with a spatially sinusoidal temperature profile created by splitting a pump laser pulse using diffractive optics and recombining the beams on the sample. A probe laser subsequently observes the in-plane transient thermal decay by measuring the intensity of light diffracted from the sample surface. In the heat diffusion regime, the shape of the transient decay curve is controlled by the in-plane thermal diffusivity \cite{johnson_phase-controlled_2012}, and thus the thermal diffusivity can be measured from the exponential time constant of the thermal decay. 

Recently, considerable interest has focused on quasiballistic heat conduction in TG, in which phonon mean free paths (MFPs) are comparable to the grating period \cite{ma_two-parameter_2014,johnson_direct_2013,maznev_onset_2011, minnich_phonon_2015,minnich_determining_2012,hua_transport_2014}. Prior works have demonstrated that observations of thermal transport in this regime yield information about the MFP spectrum of thermal phonons in a technique known as MFP spectroscopy\cite{minnich_determining_2012,maznev_onset_2011}. A number of models have been proposed to explain these results, including a two-fluid model \cite{maznev_onset_2011}, a two-parameter nondiffusive model \cite{ma_two-parameter_2014}, and related models for time-domain thermoreflectance experiments \cite{koh_nonlocal_2014,wilson_two-channel_2013, vermeersch_superdiffusive_2015, maassen_steady-state_2015}. Recently, we reported an exact analytic solution of the Boltzmann transport equation (BTE) for 1D transport in TG \cite{hua_transport_2014}.

However, often the thermal transport induced by the pump pulse is not purely 1D due to the finite optical penetration length of light in the sample \cite{johnson_phase-controlled_2012}. For example, applying TG to a GaAs wafer generates the externally defined in-plane grating but also an exponential decay in the cross-plane direction, resulting in both cross-plane and in-plane heat conduction and a thermal decay that is no longer a pure exponential \cite{luckyanova_anisotropy_2013,johnson_experimental_2011}. While in the diffusive case these measurements can be readily interpreted with Fourier's law \cite{luckyanova_anisotropy_2013}, in the quasiballistic regime interpreting the thermal decays is difficult because a unique thermal time constant can no longer be clearly identified.

In this work, we provide a theoretical framework for treating multidimensional heat conduction in TG based on an analytic Green's function for the BTE we recently reported \cite{hua_analytical_2014}. We find that the transport in 2D TG is considerably more complicated than that in 1D because the heating profile contains a spectrum of spatial frequencies rather than a single spatial frequency as in the 1D case. Nevertheless, we demonstrate a method to recover the MFP spectrum from these measurements without any fitting parameters or knowledge of the MFPs. This work extends the capability of MFP spectroscopy using TG from thin films to bulk opaque solids.

\section{Theory} \label{sec:general}

\subsection{Boltzmann Equation}
We begin by briefly reviewing the derivation of the multidimensional Green's fuction to the BTE as reported by Hua and Minnich\cite{hua_analytical_2014}. The BTE under the relaxation time approximation is given by:
\begin{equation} \label{eq:BTE_energy}
\frac{\partial g_{\k}}{\partial t}+ \textbf{v} \cdot \nabla_{\mathbf{r}}  g_{\k} = -\frac{g_{\k}-g_0(\Tzt)}{\tau_{\k}}+Q_{\k},
\end{equation}
where $g_{\k} = \hbar \omega (f_{\k}(\mathbf{r},t,\k)-f_0(T_0) )$ is the desired deviational distribution function, $Q_{\k}(\r,t)$ is the spectral volumetric heat generation, $\vzk$ is the phonon group velocity, $\tk$ is the phonon relaxation time, and $\k$ is the phonon wavevector in phase space.   $g_0(\Tzt)$ is the local equilibrium deviational distribution function and is given by:
\begin{equation}
g_0(\Tzt) = \hbar \w (f_{BE}(\Tzt) - f_0(T_0) ) \approx C_{\k} \Delta \Tzt.
\label{eq:BEDist_Linearized}
\end{equation}
assuming a small temperature rise $\Delta \Tzt = \Tzt - T_0$ relative to a reference temperature $T_{0}$. Here, $\hbar$ is the reduced Planck constant, $\w(\k)$ is the phonon frequency given by the dispersion relation, $f_{BE}(\Tzt)$ is the local Bose-Einstein distribution, $f_{0}(T_{0})$ is the Bose-Einstein distribution at $T_{0}$, and $C_{\k} = k_{B} (\chi/\sinh(\chi))^{2}$ is the mode specific heat, where $\chi=\hbar \w(\k)/2 k_{B} T_{0}$. To close the problem, energy conservation is used to relate $g_{\k}$ to $\Delta \Tzt$ as 
\begin{equation}
\sum_{\k} \left[\frac{g_{\k}}{\tk}-\frac{C_{\k}}{\tk} \Delta \Tzt \right] = 0,
\label{eq:EnergyConservation}
\end{equation}
where the sum is performed over all phonon modes in the Brillouin zone.

An analytic solution to the above equations is obtained by performing a Fourier transform over space and time in Eq.\ \ref{eq:BTE_energy}, substituting the result into Eq.\ \ref{eq:EnergyConservation}, and solving for the temperature response $\Delta \widetilde{T}(\eta,\xi_{x}, \xi_{y}, \xi_{z})$, where $\eta$ and $\xi_{x}, \xi_{y}, \xi_{z} $ are the Fourier variables in the time and spatial domains, respectively. This procedure assumes that the thermal transport occurs in an infinite domain so that a Fourier transform can be performed. The result for an isotropic crystal is:

\begin{equation}\label{eq:Temperature_FourierTransform}
\Delta \widetilde{T}(\eta,\xi_x,\xi_y,\xi_z) = \frac{\int^{\omega_m}_0\frac{\widetilde{Q}_{\k}(\eta, \xi_{x}, \xi_{x}, \xi_{z})}{\Lambda_{\omega}\xi}\text{tan}^{-1}\left(\frac{\Lambda_{\omega}\xi}{1+i\eta\tau_{\omega}}\right) d\omega}{\int^{\omega_m}_0 \frac{C_{\omega}}{\tau_{\omega}}\left[1-\frac{1}{\Lambda_{\omega}\xi}\text{tan}^{-1}\left(\frac{\Lambda_{\omega}\xi}{1+i\eta\tau_{\omega}}\right)\right]d\omega},
\end{equation}

where the magnitude of spatial frequency $\xi = \sqrt{ \xi_{x}^{2} + \xi_{y}^{2} + \xi_{z}^{2} }$. In the weakly quasiballistic regime where $\eta \tk \ll 1$, the result can be simplified by expanding all terms in powers of $\eta \tk$ and keeping only linear terms. The result is:

\begin{eqnarray}\label{eq:Temperature_FourierTransform}
\Delta \widetilde{T}(\eta,\xi_{x}, \xi_{y}, \xi_{z}) &=& \frac{ \widetilde{Q}(\eta, \xi_{x}, \xi_{x}, \xi_{z}) } { i \eta C + \kappa(\xi) \xi^{2}}
\end{eqnarray}
where $\widetilde{Q}(\eta, \xi_{x}, \xi_{x}, \xi_{z})$ is the heating profile summed over all phonon wavevectors and $\kappa(\xi)$ is the apparent thermal conductivity that depends only on $\xi$, reflecting the isotropy of the crystal. The apparent thermal conductivity is given by:

\begin{eqnarray}
\kappa(\xi) & = & \int_{0}^{\infty} f(\Lambda) S(Kn) d\Lambda \label{eq:alphaxi} \\
S(Kn) & = & \frac{3}{Kn^{2}} \left( 1 - \frac{ \tan^{-1}{Kn}}{Kn} \right) \label{eq:S}
\end{eqnarray}

where $f(\Lambda)$ is the MFP distribution, $\Lambda$ is the MFP, $S(Kn)$ is the suppression function, and the Knudsen number $Kn=\xi \Lambda$.

We now specialize this result for the 2D TG experiment. Along the in-plane, or $x$ direction, the heating profile is a sinusoid with spatial frequency $\xi_{x}$. Along the cross-plane, or $y$ direction, the heating profile is a decaying exponential with decay length $\beta_{1}$ corresponding to the optical depth for the pump beam in the solid. The probe beam samples this temperature profile with an optical depth $\beta_{2}$ corresponding to the probe wavelength. The heating along the $z$ direction is uniform and hence $\xi_{z}=0$. Recalling the expression for the Fourier transform of a decaying exponential, the temperature profile in 2D TG can thus be written as:

\begin{equation}
\label{eq:DTxi}
\Delta \widetilde{T}(\eta,\xi_{x}) =  A \int_{0}^{\infty} \frac{ (\beta_{1}^{2} + \xi_{y}^{2})^{-1} (\beta_{2}^{2} + \xi_{y}^{2})^{-1} }{i \eta C + \kappa(\xi) \xi^{2}} d\xi_{y}
\end{equation}

where $A$ is an arbitrary constant. This equation can be easily converted to the time domain as:

\begin{equation} \label{eq:DTt}
\Delta T(t, \xi_{x}) = A' \int_{0}^{\infty} \frac{\exp(-\alpha(\xi) \xi^{2} t) }{(\beta_{1}^{2} + \xi_{y}^{2}) (\beta_{2}^{2} + \xi_{y}^{2}) } d\xi_{y}
\end{equation}

where $\alpha(\xi) = \kappa(\xi)/C$ is the apparent thermal diffusivity and $A' = A/C$.

Equation \ref{eq:DTt} shows that the transient temperature profile for a 2D grating is a multiexponential decay with a spectrum of time constants determined by the apparent thermal conductivities. The relative contribution of each exponential decay to the total signal is weighted by the Fourier transform of the cross-plane heating profile. This situation is considerably more complicated than in the 1D case, for which the sole spatial frequency in the problem is externally defined by the grating period. In the 1D case, a single apparent thermal conductivity can be identified from the exponential time constant of the observed thermal decay for a specific grating period, and a set of thermal conductivities versus grating period can then be used to directly recover the MFP spectrum \cite{minnich_determining_2012}.

 In this present 2D case, however, the thermal decay is no longer a single exponential but a multiexponential. Such multiexponential decays occur often in  science, for example in time-resolved measurements of fluorescent decays \cite{enderlein_fast_1997}, and are challenging to interpret because recovering all the exponential time constants from a multiexponential decay is an ill-posed problem for which a unique solution does not exist. Thus in the 2D case, an inverse problem must be solved to recover the apparent thermal conductivities from the measured temperature decays before any further information about the thermal phonons can be extracted. Further, the inverse problem is nonlinear because the function to be recovered is inside an exponential.

\subsection{Optimization Problem}
This inverse problem can be solved using constrained nonlinear optimization methods, thereby yielding the apparent thermal conductivities as a function of spatial frequency $\xi$. These values can then be directly used as input to the original inverse problem of recovering the MFP spectrum described in Ref.\ \cite{minnich_determining_2012} that is readily solved using convex optimization.

The present inverse problem is to recover the function $\alpha(\xi)$ from a set of transient thermal decays, $\Delta T(t, \xi_{x})$, where the spatial frequency along the in-plane direction $\xi_{x}$ can be externally controlled by varying the grating period. The optical decay lengths $\beta_{1}$ and $\beta_{2}$ are assumed to be fixed by the material and optical wavelengths used in the experiment, thereby specifying the spatial frequencies along the cross-plane direction, $\xi_{y}$. This prototypical problem has been addressed using a number of methods in the past, including Prony's method \cite{marple_digital_1987}, Fourier methods \cite{provencher_fourier_1976}, and linearization to enable linear optimization algorithms\cite{novikov_linear_1999,enderlein_fast_1997,curtis_analysis_1970}. In this work, we use nonlinear optimization methods to solve the problem.

Let us discretize this problem by assuming with have $N$ transient data sets $\Delta T_{i}(t, \xi_{x,i})$ corresponding to different grating periods used in the experiment. Our goal is to adjust $M$ apparent thermal diffusivities $\alpha_{j}$ that depend on spatial frequency to match Eq.\ \ref{eq:DTt} to the experimental data set. This problem can be classified as a nonlinear constrained optimization by defining a cost function that measures the difference between the simulated and actual thermal decays and attempting to minimize this cost function by varying $\alpha_{j}$. As with many inverse problems, attempts to solve this problem without any constraints will likely not succeed because a unique solution does not exist. The desired solution can only be recovered from the measurements if known physical facts about the solution are incorporated into the optimization. 

Fortunately, a number of facts are known about the solution. Based on the behavior of the derivatives of the suppression function, one can derive that the first derivative of $\alpha(\xi)$ must be negative, the second derivative must be positive, and the third derivative must be negative. These constraints can be implemented by approximating the derivative with finite differences and incorporating the resulting equations into the optimization as linear inequality constraints.

Second, the solution cannot have jumps between adjacent $\alpha_{j}$ that are too large. Considering that the spatial frequencies $\xi_{j}$ are uniformly spaced, the maximum difference between $\alpha_{j+1}$ and $\alpha_{j}$ is:

\begin{equation} \label{}
\alpha_{j} - \alpha_{j+1} =   \int_{0}^{\infty} \alpha(\Lambda) \left( - \frac{dS}{d Kn} \right) \Lambda d\Lambda  \Delta \xi \lesssim   \alpha_{bulk} \Lambda_{med} \Delta \xi
\end{equation}
where $ \alpha(\Lambda) = \kappa(\Lambda)/C$ is the differential thermal diffusivity distribution versus MFP and $\Lambda_{med}$ is an estimate of the median MFP of the material. This constraint can also be incorporated into the optimization as a linear inequality constraint.

Third, the apparent thermal diffusivities do not vary substantially over the range of experimentally attainable grating periods, and thus upper and lower bounds on the allowed values of $\alpha_{j}$ can be imposed. For instance, for the analysis performed for Si in the next section, the apparent thermal diffusivities must be between $1$ and $100$ mm$^{2}$/s. Similar bounds can be imposed for other solids for which basic knowledge of bulk thermal properties is known.

With these constraints, the optimization problem can be solved with standard optimization routines in MATLAB. We use the \textit{fmincon} function in MATLAB for this work. The cost function to be minimized is defined as the magnitude of the difference between the calculated and measured temperature decays for $N$ different grating periods $\xi_{x}$. The constraints described above are put into matrix form and included in the optimization as linear inequality constraints. The outputs of the optimization are $M$ discrete values of the function $\alpha(\xi_{i})$ that minimize the cost function while satisfying the constraints.


\section{Results} \label{sec:results}

\begin{figure}
\begin{center}
\includegraphics[width=1\textwidth]{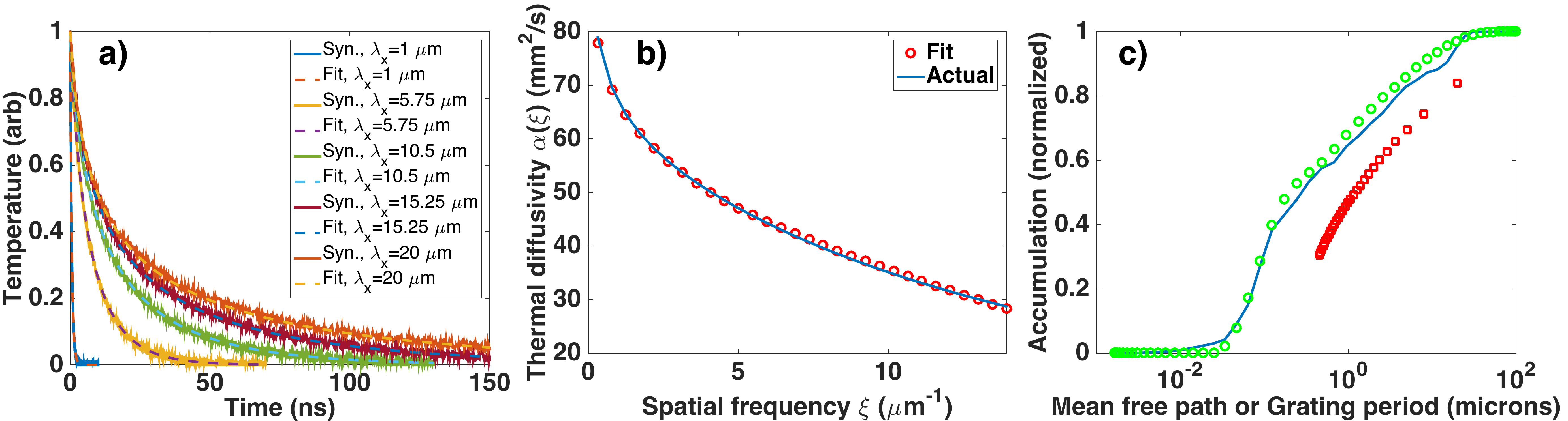}
\caption{(a) Synthesized temperature decay curves (solid lines) and decay curves corresponding to the thermal diffusivity function reconstructed by nonlinear optimization (dashed lines). The reconstructed thermal decays are in good agreement with the original synthesized curves. (b) Reconstructed thermal diffusivity $\alpha(\xi)$ (symbols) with equality constraints compared to the actual function (solid line). The reconstructed function is in good agreement with the actual function. (c) Reconstructed MFP spectrum using the thermal diffusivity result from (b). The reconstructed MFP spectrum and the actual distribution agree well with each other.}
\label{fig:DT}
\end{center}
\end{figure}

To demonstrate the reconstruction, we first synthesize temperature decay curves $\Delta T_{i}$ for different grating periods $\lambda_{i}=2\pi/\xi_{x,i}$ between 500 nm to 20 microns. The decay curves are obtained by calculating the apparent thermal diffusivities versus spatial frequency using Eq.\ \ref{eq:alphaxi} and the phonon dispersion and lifetimes for Si at room temperature as calculated by Jes$\acute{\text{u}}$s Carrete and Natalio Mingo using ShengBTE \cite{Mingo2012, Mingo2014} and Phonopy\cite{Phonony} from interatomic force constants obtained with VASP.\cite{Hafner1993, Hafner1994, Furthmuller1996a, Furthmuller1996b} These results are then inserted into Eq.\ \ref{eq:DTt}, yielding transient decay curves for a particular grating period. 

The resulting decay curves, with random noise added to simulate experimental uncertainty, are shown in Fig.\ \ref{fig:DT}a. We then use the optimization procedure described in the previous section to reconstruct the apparent thermal diffusivities. For this problem, we discretized the spatial frequencies $\xi_{i}$ into $M=30$ equally spaced points, and attempted to fit $N=25$ temperature decay curves. The extinction coefficients in the cross-plane direction are taken to be $\beta_{1} = \beta_{2} = 1/500$ nm$^{-1}$.

In the first set of results, we took the minimum grating period to be 500 nm, a smaller value than is typically achieved experimentally but still possible with commercially available optics. The availability of data at a grating period that is comparable to the optical decay length is very helpful because the thermal decay corresponds predominantly to the single spatial frequency defined by the imposed grating rather than a spectrum of spatial frequencies as in the general case. As shown in Fig.\ \ref{fig:DT}b, the reconstructed thermal diffusivity is in excellent agreement with the actual thermal diffusivity despite the presence of noise in the original temperature decay curves in Fig.\ \ref{fig:DT}a.

Using this reconstructed thermal diffusivity function, we then close the problem by performing the convex optimization procedure described in Ref.\ \cite{minnich_determining_2012} to recover the desired MFP spectrum. This result is given in Fig.\ \ref{fig:DT}c, and shows that the reconstructed distribution is in good agreement with the actual MFP spectrum. Our approach of solving two inverse problems, starting from the original transient decay curves, is thus successfully able to reconstruct the desired MFP spectrum without any knowledge of the MFPs or thermal properties of the material.

\begin{figure}
\begin{center}
\includegraphics[width=0.95\textwidth]{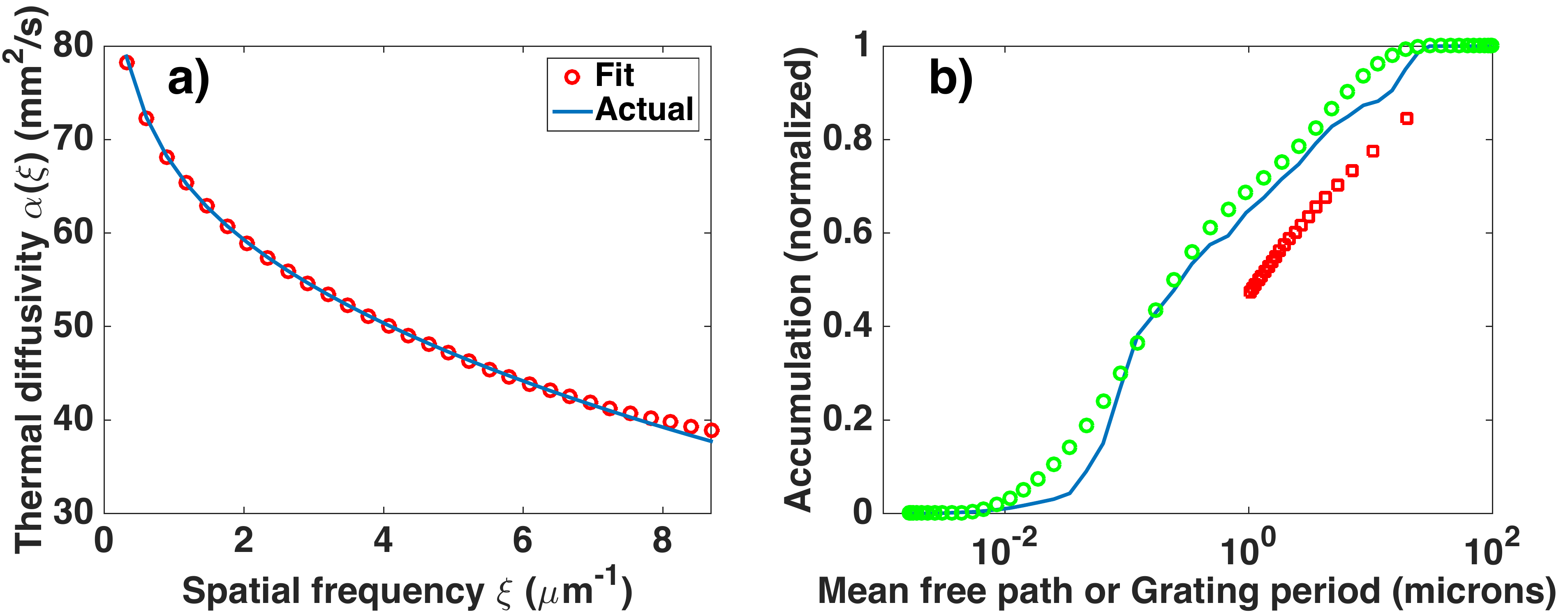}
\caption{(a) Reconstructed thermal diffusivity with data from grating periods between 1-20 microns (symbols) compared to the actual function (solid line). The reconstruction is good but some discrepancy exists at higher spatial frequencies. (b) Reconstructed MFP spectrum (green circles) using the only the thermal diffusivities in (a) with spatial frequency less than $2\pi$ $\um^{-1}$ (red squares) compared to the actual spectrum (solid line). The x-axis corresponds to MFP for the MFP spectrum or the grating period for thermal diffusivity data. The reconstructed spectrum agrees well with the actual function provided high spatial frequency data points from (a) are discarded.}
\label{fig:1micron}
\end{center}
\end{figure}

The reconstruction without data at grating periods comparable to the optical decay length is more challenging but still feasible. The reconstructed thermal diffusivity versus spatial frequency obtained when only using grating periods from 1-20 microns is shown in Fig.\ \ref{fig:1micron}a. The reconstruction and correct answer are in generally good agreement, though a discrepancy exists at high spatial frequencies. Performing the second inverse problem with this input results in a poor reconstruction of the MFP spectrum for values of the MFP below 1 micron (not shown). To circumvent this limitation, we recognize that the reconstructed values for spatial frequencies larger than that corresponding to the minimum grating period, $\approx 2 \pi/1$ $\um^{-1}$, are only minimally constrained by the available data, and may thus be unreliable. Discarding these points, we perform the reconstruction with the subset of data corresponding to spatial frequencies less than $\approx 2 \pi$ $\um^{-1}$. The result is given in Fig.\ \ref{fig:1micron}b, demonstrating quite reasonable agreement with the actual MFP spectrum. Thus, the MFP spectrum can be effectively extracted from 2D TG measurements even if multidimensional heat conduction occurs in all of the measured data. The reconstruction can be made more robust if data with grating period smaller than the optical decay length can be obtained so that the heat conduction is primarily one-dimensional for some of the measurements.

\section{Anisotropy of apparent thermal conductivity}

A recent experimental work reported the anisotropic failure of Fourier's law in a time-domain thermoreflectance (TDTR) experiment in which heat conduction occurs both in- and cross-plane \cite{wilson_anisotropic_2014}. The authors observed that the in-plane thermal conductivity of Si appeared to be nearly a factor of two smaller than the cross-plane value when the pump beam diameter was on the order of a few microns, comparable to the phonon MFPs in Si.

Although the samples in the TDTR experiment consist of a film on top of a substrate rather than only a substrate as in TG, we can qualitatively examine this report using our multidimensional solution of the BTE to describe heat conduction in the substrate. First, consider the form of apparent thermal conductivity expressed in Eq.\ \ref{eq:Temperature_FourierTransform}. This equation shows that for an isotropic crystal, the apparent thermal conductivity depends only on the magnitude of spatial frequency rather than the components along each direction, implying that the apparent thermal conductivity must be isotropic in real space. Since Si is a cubic crystal that is thermally isotropic, our BTE solution indicates that the apparent thermal conductivity must be isotropic even in the quasiballistic regime regardless of the anisotropy of the thermal gradients.

The obvious question is then why an anisotropic thermal conductivity appears to explain the data even though the apparent thermal conductivity must be isotropic. A possible answer can be found in the similarity of the Fourier's law solution used to fit the data and the exact solution given by the BTE. Recall from Eq.\ \ref{eq:DTt} that the temperature decay of a 2D grating consists of a multiexponential decay with time constants determined by the apparent thermal conductivities that depend only on the magnitude of the spatial frequency. On the other hand, the equation giving the Fourier's law solution for an anisotropic thermal conductivity is given by:

\begin{equation} \label{eq:AniFourier}
\Delta T(t, \xi_{x}) = A' \int_{0}^{\infty} \frac{\exp[-(\alpha_{x} \xi_{x}^{2} + \alpha_{y} \xi_{y}^{2}) t] }{(\beta_{1}^{2} + \xi_{y}^{2}) (\beta_{2}^{2} + \xi_{y}^{2}) } d\xi_{y}
\end{equation}

where $\alpha_{x}$ and $\alpha_{y}$ are the thermal diffusivities along the $x$ and $y$ directions, respectively, and are used as fitting parameters to fit the measured thermal decay. Consider the similarities between Eqs.\ \ref{eq:DTt} and \ref{eq:AniFourier}. In Eq.\ \ref{eq:DTt}, the integral contains a spectrum of thermal diffusivities that continuously vary with spatial frequency $\xi$. In Eq.\ \ref{eq:AniFourier}, there are only two thermal diffusivities but these separately multiply $\xi_{x}$ and $\xi_{y}$. Due to the averaging caused by the integral over $\xi_{y}$, it is possible to reproduce a transient thermal decay by appropriately adjusting $\alpha_{x}$ and $\alpha_{y}$ even though this anisotropic solution is inconsistent with the actual BTE solution.

\begin{figure}
\begin{center}
\includegraphics[width=0.8\textwidth]{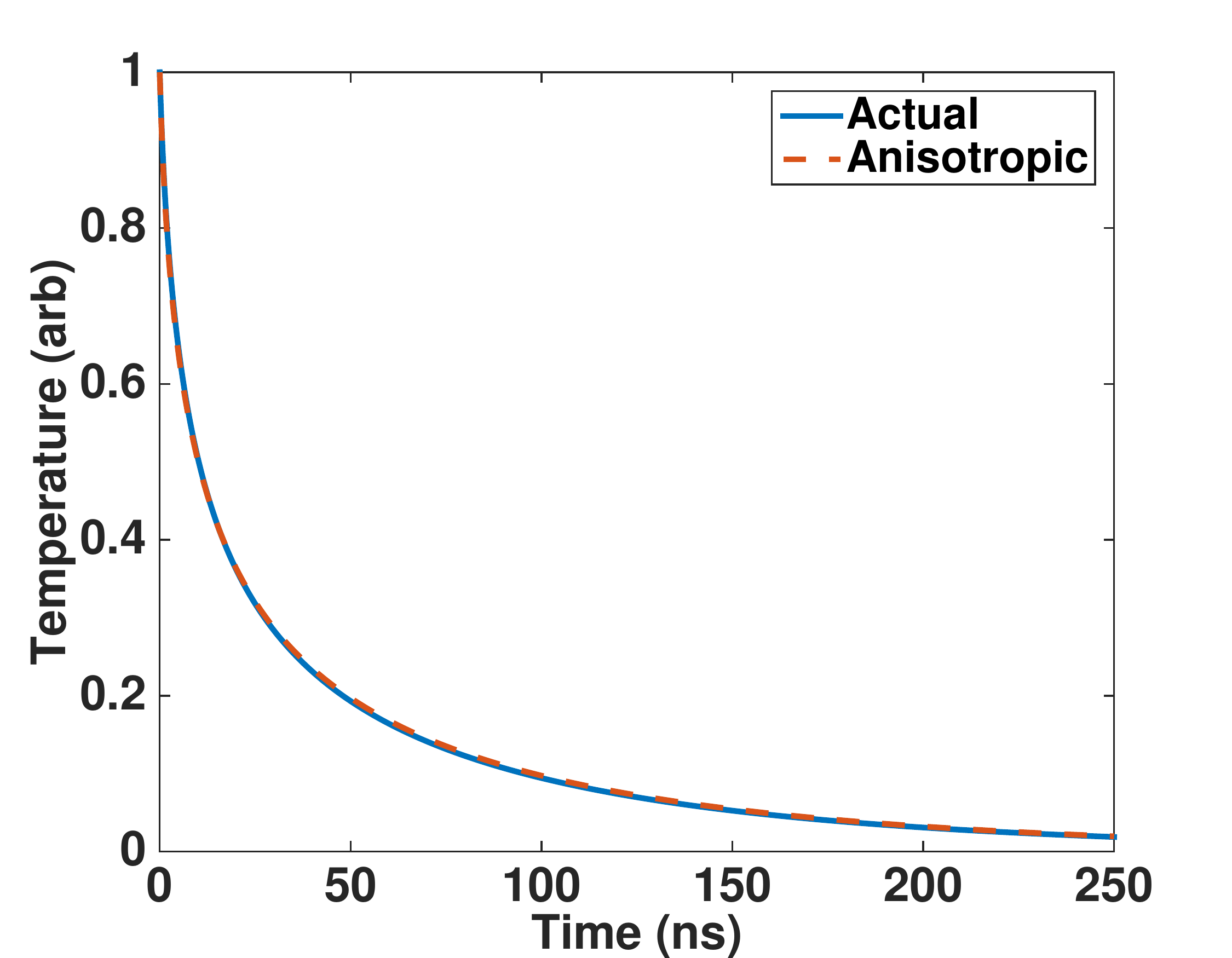}
\caption{Calculated temperature versus time from the BTE solution (solid line) and from an anisotropic Fourier model (dashed line). The anisotropic Fourier model is able to explain the observed thermal decay even though the apparent thermal conductivity is isotropic.}
\label{fig:ani}
\end{center}
\end{figure}

As an example, we calculate the thermal decay derived from the BTE using Eq.\ \ref{eq:DTt}, then attempt to fit this decay curve using Eq.\ \ref{eq:AniFourier} with $\alpha_{x}$ and $\alpha_{y}$ as fitting parameters. For $\lambda=20$ $\um$ and $\beta = 1/500$ nm$^{-1}$, we are able to fit the thermal decay nearly exactly using $\alpha_{x} = 0.85 \alpha_{bulk}$ and $\alpha_{y}/\alpha_{x}=0.86$ as shown in Fig.\ \ref{fig:ani}. However, this successful fitting does not imply that the apparent thermal conductivity is anisotropic, but rather highlights that multiple solutions can explain the same data for this ill-posed problem. This observation demonstrates the utility of having the exact analytical solution to the multidimensional BTE as presented here.  

Our result shows that anisotropic thermal gradients in bulk isotropic crystals are not sufficient to yield apparent thermal conductivities that are anisotropic. It is possible that an anisotropic heating profile, in combination with the metal-semiconductor interface present in TDTR, could lead to anisotropic apparent thermal conductivity; further study with a BTE solution that incorporates the interface is necessary to completely resolve the issue.

\section{Summary}

We have analyzed multidimensional quasiballistic thermal transport in TG using an exact analytic solution of the BTE. We find that the thermal decay in multidimensional TG consists of a multiexponential decay with the time constants determined by the apparent thermal conductivities as a function of spatial frequency. By solving a nonlinear inverse problem, we are able to recover these apparent thermal conductivities and thereby obtain the MFP spectrum of the thermal phonons without any fitting parameters or knowledge of the MFPs. This work extends the MFP spectroscopy method with TG to opaque bulk materials.

\section{Acknowledgements}

The author thanks Jes$\acute{\text{u}}$s Carrete and Natalio Mingo for providing the first-principles data for Si. This work was sponsored in part by the National Science Foundation under Grant no. CBET CAREER 1254213, and by Boeing under the Boeing-Caltech Strategic Research \& Development Relationship Agreement.

\bibliographystyle{is-unsrt}
\bibliography{2DTG}

\begin{thebibliography}{10}
\ifx \showCODEN  \undefined \def \showCODEN #1{CODEN #1}  \fi
\ifx \showISBN   \undefined \def \showISBN  #1{ISBN #1}   \fi
\ifx \showISSN   \undefined \def \showISSN  #1{ISSN #1}   \fi
\ifx \showLCCN   \undefined \def \showLCCN  #1{LCCN #1}   \fi
\ifx \showPRICE  \undefined \def \showPRICE #1{#1}        \fi
\ifx \showURL    \undefined \def \showURL {URL }          \fi
\ifx \path       \undefined \input path.sty               \fi
\ifx \ifshowURL \undefined
     \newif \ifshowURL
     \showURLtrue
\fi

\bibitem{maznev_optical_1998}
A.~A. Maznev, K.~A. Nelson, and J.A. Rogers.
\newblock Optical heterodyne detection of laser-induced gratings.
\newblock {\em Optics Letters}, 23\penalty0 (16):\penalty0 1319--1321, 1998.
\newblock \ifshowURL {\showURL
  \path|http://ol.osa.org/abstract.cfm?URI=ol-23-16-1319|}\fi.

\bibitem{johnson_phase-controlled_2012}
Jeremy~A. Johnson, Alexei~A. Maznev, Mayank~T. Bulsara, Eugene~A. Fitzgerald,
  T.~C. Harman, S.~Calawa, C.~J. Vineis, G.~Turner, and Keith~A. Nelson.
\newblock Phase-controlled, heterodyne laser-induced transient grating
  measurements of thermal transport properties in opaque material.
\newblock {\em Journal of Applied Physics}, 111\penalty0 (2):\penalty0
  023503--023503--7, January 2012.
\newblock \showISSN{00218979}.
\newblock \ifshowURL {\showURL
  \path|http://jap.aip.org/resource/1/japiau/v111/i2/p023503_s1|}\fi.

\bibitem{ma_two-parameter_2014}
Yanbao Ma.
\newblock A two-parameter nondiffusive heat conduction model for data analysis
  in pump-probe experiments.
\newblock {\em Journal of Applied Physics}, 116\penalty0 (24):\penalty0 243505,
  December 2014.
\newblock \showISSN{0021-8979, 1089-7550}.
\newblock \ifshowURL {\showURL
  \path|http://scitation.aip.org/content/aip/journal/jap/116/24/10.1063/1.4904355|}\fi.

\bibitem{johnson_direct_2013}
Jeremy~A. Johnson, A.~A. Maznev, John Cuffe, Jeffrey~K. Eliason, Austin~J.
  Minnich, Timothy Kehoe, Clivia M.~Sotomayor Torres, Gang Chen, and Keith~A.
  Nelson.
\newblock Direct {Measurement} of {Room}-{Temperature} {Nondiffusive} {Thermal}
  {Transport} {Over} {Micron} {Distances} in a {Silicon} {Membrane}.
\newblock {\em Physical Review Letters}, 110\penalty0 (2):\penalty0 025901,
  January 2013.
\newblock \ifshowURL {\showURL
  \path|http://link.aps.org/doi/10.1103/PhysRevLett.110.025901|}\fi.

\bibitem{maznev_onset_2011}
A.~A. Maznev, Jeremy~A. Johnson, and Keith~A. Nelson.
\newblock Onset of nondiffusive phonon transport in transient thermal grating
  decay.
\newblock {\em Physical Review B}, 84\penalty0 (19):\penalty0 195206, November
  2011.
\newblock \ifshowURL {\showURL
  \path|http://link.aps.org/doi/10.1103/PhysRevB.84.195206|}\fi.

\bibitem{minnich_phonon_2015}
A.~J. Minnich.
\newblock Phonon heat conduction in layered anisotropic crystals.
\newblock {\em Physical Review B}, 91\penalty0 (8):\penalty0 085206, February
  2015.
\newblock \ifshowURL {\showURL
  \path|http://link.aps.org/doi/10.1103/PhysRevB.91.085206|}\fi.

\bibitem{minnich_determining_2012}
A.~J. Minnich.
\newblock Determining {Phonon} {Mean} {Free} {Paths} from {Observations} of
  {Quasiballistic} {Thermal} {Transport}.
\newblock {\em Physical Review Letters}, 109\penalty0 (20):\penalty0 205901,
  November 2012.
\newblock \ifshowURL {\showURL
  \path|http://link.aps.org/doi/10.1103/PhysRevLett.109.205901|}\fi.

\bibitem{hua_transport_2014}
Chengyun Hua and Austin~J. Minnich.
\newblock Transport regimes in quasiballistic heat conduction.
\newblock {\em Physical Review B}, 89\penalty0 (9):\penalty0 094302, March
  2014.
\newblock \ifshowURL {\showURL
  \path|http://link.aps.org/doi/10.1103/PhysRevB.89.094302|}\fi.

\bibitem{koh_nonlocal_2014}
Yee~Kan Koh, David~G. Cahill, and Bo~Sun.
\newblock Nonlocal theory for heat transport at high frequencies.
\newblock {\em Physical Review B}, 90\penalty0 (20):\penalty0 205412, November
  2014.
\newblock \ifshowURL {\showURL
  \path|http://link.aps.org/doi/10.1103/PhysRevB.90.205412|}\fi.

\bibitem{wilson_two-channel_2013}
R.~B. Wilson, Joseph~P. Feser, Gregory~T. Hohensee, and David~G. Cahill.
\newblock Two-channel model for nonequilibrium thermal transport in pump-probe
  experiments.
\newblock {\em Physical Review B}, 88\penalty0 (14):\penalty0 144305, October
  2013.
\newblock \ifshowURL {\showURL
  \path|http://link.aps.org/doi/10.1103/PhysRevB.88.144305|}\fi.

\bibitem{vermeersch_superdiffusive_2015}
Bjorn Vermeersch, Jes{\'u}s Carrete, Natalio Mingo, and Ali Shakouri.
\newblock Superdiffusive heat conduction in semiconductor alloys. {I}.
  {Theoretical} foundations.
\newblock {\em Physical Review B}, 91\penalty0 (8):\penalty0 085202, February
  2015.
\newblock \ifshowURL {\showURL
  \path|http://link.aps.org/doi/10.1103/PhysRevB.91.085202|}\fi.

\bibitem{maassen_steady-state_2015}
Jesse Maassen and Mark Lundstrom.
\newblock Steady-state heat transport: {Ballistic}-to-diffusive with
  {Fourier}'s law.
\newblock {\em Journal of Applied Physics}, 117\penalty0 (3):\penalty0 035104,
  January 2015.
\newblock \showISSN{0021-8979, 1089-7550}.
\newblock \ifshowURL {\showURL
  \path|http://scitation.aip.org.clsproxy.library.caltech.edu/content/aip/journal/jap/117/3/10.1063/1.4905590|}\fi.

\bibitem{luckyanova_anisotropy_2013}
Maria~N. Luckyanova, Jeremy~A. Johnson, A.~A. Maznev, Jivtesh Garg, Adam Jandl,
  Mayank~T. Bulsara, Eugene~A. Fitzgerald, Keith~Adam Nelson, and Gang Chen.
\newblock Anisotropy of the thermal conductivity in {GaAs}/{AlAs}
  superlattices.
\newblock {\em Nano Letters}, August 2013.
\newblock \showISSN{1530-6984}.
\newblock \ifshowURL {\showURL \path|http://dx.doi.org/10.1021/nl4001162|}\fi.

\bibitem{johnson_experimental_2011}
Jeremy~A. Johnson, Alexei~A. Maznev, Jeffrey~K. Eliason, Austin Minnich,
  Kimberlee Collins, Gang Chen, John Cuffe, Timothy Kehoe, Clivia~M.
  Sotomayor~Torres, and Keith~A. Nelson.
\newblock Experimental {Evidence} of {Non}-{Diffusive} {Thermal} {Transport} in
  {Si} and {GaAs}.
\newblock {\em MRS Online Proceedings Library}, 1347:\penalty0 null, 2011.

\bibitem{hua_analytical_2014}
Chengyun Hua and Austin~J. Minnich.
\newblock Analytical {Green}'s function of the multidimensional
  frequency-dependent phonon {Boltzmann} equation.
\newblock {\em Physical Review B}, 90\penalty0 (21):\penalty0 214306, December
  2014.
\newblock \ifshowURL {\showURL
  \path|http://link.aps.org/doi/10.1103/PhysRevB.90.214306|}\fi.

\bibitem{enderlein_fast_1997}
J{\"o}rg Enderlein and Rainer Erdmann.
\newblock Fast fitting of multi-exponential decay curves.
\newblock {\em Optics Communications}, 134\penalty0 (1--6):\penalty0 371--378,
  January 1997.
\newblock \ifshowURL {\showURL
  \path|http://www.sciencedirect.com/science/article/pii/S0030401896003847|}\fi.

\bibitem{marple_digital_1987}
S.~Lawrence Marple.
\newblock Digital spectral analysis with applications.
\newblock {\em Englewood Cliffs, NJ, Prentice-Hall, Inc., 1987, 512 p.}, -1,
  1987.
\newblock \ifshowURL {\showURL
  \path|http://adsabs.harvard.edu/abs/1987ph...book.....M|}\fi.

\bibitem{provencher_fourier_1976}
S~W Provencher.
\newblock A {Fourier} method for the analysis of exponential decay curves.
\newblock {\em Biophysical Journal}, 16\penalty0 (1):\penalty0 27--41, January
  1976.
\newblock \showISSN{0006-3495}.
\newblock \ifshowURL {\showURL
  \path|http://www.ncbi.nlm.nih.gov/pmc/articles/PMC1334811/|}\fi.

\bibitem{novikov_linear_1999}
Eugene~G Novikov, Arie van Hoek, Antonie J. W.~G Visser, and Johannes~W
  Hofstraat.
\newblock Linear algorithms for stretched exponential decay analysis.
\newblock {\em Optics Communications}, 166\penalty0 (1--6):\penalty0 189--198,
  August 1999.
\newblock \ifshowURL {\showURL
  \path|http://www.sciencedirect.com/science/article/pii/S003040189900262X|}\fi.

\bibitem{curtis_analysis_1970}
L.~J. Curtis, H.~G. Berry, and J.~Bromander.
\newblock Analysis of {Multi}-exponential {Decay} {Curves}.
\newblock {\em Physica Scripta}, 2\penalty0 (4-5):\penalty0 216, October 1970.
\newblock \showISSN{1402-4896}.
\newblock \ifshowURL {\showURL
  \path|http://iopscience.iop.org/1402-4896/2/4-5/015|}\fi.

\bibitem{Mingo2012}
Wu~Li, Natalio Mingo, Lucas Lindsay, David~A. Broido, Derek~A. Stewart, and
  Nebil~A. Katcho.
\newblock Thermal conductivity of diamond nanowires from first principles.
\newblock {\em Phys. Rev. B}, 85:\penalty0 195436, 2012.

\bibitem{Mingo2014}
W.~Li, J.~Carrete, NA~Katcho, and N.~Mingo.
\newblock Shengbte:a solver of the botlzmann transport equation for phonons.
\newblock {\em Computer Physics Communications}, 185:\penalty0 1747--58, 2014.

\bibitem{Phonony}
Atsushi Togo.
\newblock {\em Phonony}, v1.8.5.
\newblock \ifshowURL {\showURL \path|http://phonopy.sourceforge.net|}\fi.

\bibitem{Hafner1993}
G.~Kresse and J.~Hafner.
\newblock Ab initio molecular dynamics for liquid metals.
\newblock {\em Phys. Rev. B}, 48:\penalty0 13115--13118, Nov 1993.
\newblock \ifshowURL {\showURL
  \path|http://link.aps.org/doi/10.1103/PhysRevB.48.13115|}\fi.

\bibitem{Hafner1994}
G.~Kresse and J.~Hafner.
\newblock Ab initio molecular-dynamics simulation of the
  liquidmetalamorphous-semiconductor transition in germanium.
\newblock {\em Phys. Rev. B}, 49:\penalty0 14251--14269, May 1994.
\newblock \ifshowURL {\showURL
  \path|http://link.aps.org/doi/10.1103/PhysRevB.49.14251|}\fi.

\bibitem{Furthmuller1996a}
Kresse G. and J~Furthmuller.
\newblock Efficiency of ab-initio total energy calculations for metals and
  semiconductors using a plane-wave basis set.
\newblock {\em Computational Materials Science}, 6:\penalty0 15--50, 1996.

\bibitem{Furthmuller1996b}
G.~Kresse and J.~Furthmuller.
\newblock Efficient iterative schemes for ab initio total-energy calculations
  using a plane-wave basis set.
\newblock {\em Phys. Rev. B}, 54:\penalty0 11169--11186, Oct 1996.
\newblock \ifshowURL {\showURL
  \path|http://link.aps.org/doi/10.1103/PhysRevB.54.11169|}\fi.

\bibitem{wilson_anisotropic_2014}
R.~B. Wilson and David~G. Cahill.
\newblock Anisotropic failure of {Fourier} theory in time-domain
  thermoreflectance experiments.
\newblock {\em Nature Communications}, 5, October 2014.
\newblock \ifshowURL {\showURL
  \path|http://www.nature.com.clsproxy.library.caltech.edu/ncomms/2014/141001/ncomms6075/full/ncomms6075.html|}\fi.

\end{thebibliography}

\end{document}